\documentclass[preprint,12pt]{elsarticle}

\usepackage{amsfonts,amsmath,amssymb,bm,a4wide,graphicx,makeidx,multicol,color,feynmp}
\usepackage{pst-plot,pstricks-add}
\biboptions{sort&compress}
\journal{Physics Letters B}
\newcommand{\bs}{\boldsymbol}
\newcommand{\rme}{\mathrm{e}}
\newcommand{\rmi}{\mathrm{i}}
\newcommand{\be}{\begin{equation}}
\newcommand{\ee}{\end{equation}}
\begin{document}
\begin{frontmatter}
\title{Spin light of electron in dense neutrino fluxes}

\author{
Ilya Balantsev$^{a}$\footnote{balantsev@physics.msu.ru},
Alexander Studenikin$^{b,c}$\footnote{studenik@srd.sinp.msu.ru}
}

\address{

    $^{a}$Sternberg Astronomical Institute, \ Moscow State University,
        119991 Moscow, Russia

    $^{b}$Faculty of Physics, Department of Theoretical Physics, \ Lomonosov Moscow State University,
        119991 Moscow, Russia

    $^{c}$Joint Institute for Nuclear Research, 141980 Dubna, Russia
}

\begin{abstract}
An electron motion in a dense neutrino flux  is investigated. The Dirac equation exact solutions for
the electron energy  and wave function in this external environment are obtained. On this basis we predict the existence of an electromagnetic radiation that can be emitted by electrons in dense neutrino fluxes. We term this phenomenon ``the spin light of electron in dense neutrino flux" ($SLe_{\nu}$). The main properties of the $SLe_{\nu}$ are studied. We argue that the $SLe_{\nu}$ in dense neutrino fluxes should have important consequences in astrophysics and for  supernovae processes in particular.
\end{abstract}

\end{frontmatter}

\section{Introduction}

The problem of neutrino propagation and interactions in dense external environments
is one of the most important issues
 of the present-day particle physics. Neutrinos play an important role in very complex and physically diverse processes as of core-collapse supernovae that provides fascinating playground for new or exotic phenomena.
 A detailed review on supernova mechanisms is given in \cite{Bethe:1990mw}, a discussion on some of the progress that has been made recently in the simulations of stellar core-collapse and supernova explosions can be also found in \cite{Janka:2006fh}.

The physical processes that lead from stellar core-collapse to supernova explosion provide
an extremely important and interesting astrophysical application of the discussed problem of neutrino interactions in external environments.
  Under the influence of the extreme external conditions, such as dense matter and strong magnetic fields, neutrinos can manifest their yet hidden fundamental properties that might open a window to new physics.

Recently studies of neutrino electromagnetic properties (see \cite{Giunti:2014ixa, Broggini:2012df,Giunti:2008ve} for a review)  have revealed a new mechanism of electromagnetic radiation that can be emitted by
a neutrino propagating in dense matter termed the spin
light of neutrino in matter ($SL\nu$) \cite{Lobanov:2002ur}. At first the $SL\nu$ was investigated within the quasi-classical treatment (see also \cite{Lobanov:2004um,Grigoriev:2004bm}) based on use of the generalized Bargmann-Michel-Telegdi equation for the neutrino spin evolution in the background environment \cite{Lobanov:2001ar,Dvornikov:2002rs}.

The quantum theory of the $SL\nu$ was first revealed in our studies
\cite{Studenikin:2004dx,Grigorev:2005sw} ( see also \cite{Lobanov:2005zn}) within implication of  the so called ``method of wave equations exact solutions" that implies use of exact solutions of modified Dirac equations that contain the corresponding effective potentials accounting for the matter influence on neutrinos \cite{Studenikin:2004dx,Grigorev:2005sw,Studenikin:2005bq,Studenikin:2007zza,Studenikin:2008qk,Balantsev:2010zw,
Balantsev:2013aya,Studenikin:2012vi}.
Different aspects related to the proposed the $SL\nu$ have been discussed and investigated recently. For instance, the importance of the plasma influence on the proposed new mechanism of electromagnetic radiation as well as corrections to the effective matter density due to nonlocality of neutrino interaction with particles of the environment have been considered in the subsequent series of papers dedicated to the $SL\nu$ (see \cite{Kuznetsov:2007ar,Grigoriev:2012pw,Kuznetsov:2013tta}). The $SL\nu$ mechanism was also considered for the case when in addition to matter  a gravitational field is also present \cite{Grigoriev:2004bm}, or the model describing neutrino interactions with the environment permits of the Lorentz and CPT invariance violation \cite{Kruglov:2013oia}.

The main properties of the $SL\nu$ are summarized  \cite{Studenikin:2008qk}
in the following way:     1) a neutrino with nonzero magnetic moment when
moving in dense matter can emit electromagnetic waves; 2) the
$SL\nu$ radiation rate and  power depend on the neutrino magnetic
moment and energy, and also on the matter density;
3) for a wide range of matter densities the radiation
is beamed along the neutrino momentum;
4) the
emitted photon energy is also essentially dependent on the
neutrino energy and matter density;
5) in the most
interesting for possible astrophysical and cosmology applications
case of ultra-high energy neutrinos, the average energy of the
$SL\nu$ photons equals a reasonable part of the initial neutrino energy so that the $SL\nu$ spectrum can span up to gamma-rays.

In spite of the listed above notable properties of the $SL\nu$ its possible role and impact in astrophysical processes is constraint due to the fact that the rate of the process is proportional to the neutrino magnetic moment squared that is in fact a very small quantity for the most of theories  beyond the Standard Model (see, for instance, \cite{Giunti:2014ixa}). Other constraints on possible visualization of the $SL\nu$ are imposed by the mentioned above effects of the background plasma.

To avoid the suppression of the radiation produced by the spin light mechanism we
considered the electromagnetic radiation by an electron moving in matter (the ``spin light of electron in matter", $SLe$). From the order-of-magnitude estimation \cite{Studenikin:2005bq}, we
predicted that the ratio of rates of the $SLe$ and the $SL\nu$ in
matter is
\begin{equation}
R=\frac {\Gamma_{SLe}}{\Gamma_{SL\nu}}\sim \frac {e^2}{\omega^2
\mu^2},
\end{equation}
that gives $R \sim 10^{20}\div 10^{14}$ for the value of the neutrino magnetic
moment $\mu \sim 10^{-11}\mu_0$ and the predicted  wide range $SLe$ photon's energies $\omega \sim 5 \ MeV \div 5 \ GeV   $.
This estimation was confirmed by the direct evaluation \cite{Grigoriev:2006rv} of the $SLe$ properties based on the exact solutions of the modified Dirac equation for the electron in matter. Thus, we expect that in certain
cases the $SLe_{\nu}$ in matter would be more effective than the $SL\nu$. However, the possibility of phenomenological consequences of the $SLe$ in astrophysics  is quite not obvious.

In this Letter we continue studies of the spin light mechanism of electromagnetic radiation in a dense environment and consider a new possible realization of this mechanism. The predicted new mechanism implies the electromagnetic radiation emitted by an electron in a dense flux of ultra-relativistic neutrinos. We term this mechanism ``the spin light of electron in dense neutrino fluxes", $SLe_{\nu}$. This new realization of the spin light provides a possibility to avoid two suppression factors in the radiation rate and power, the discussed above suppression due to smallness of a neutrino magnetic moment and one due to the plasma effects.  We predict that this new realization of the spin light mechanism can have visible consequences for different astrophysical settings, for stellar core-collapse and supernova explosion phenomenology in particular.

\section{Modified Dirac equation}

Consider a neutrino flux propagating through medium composed of different particles like electrons, protons and neutrons. This situation can be found in various astrophysical settings, for example, in supernovae. In our approach  we suppose that the neutrino flux is described by a set of macroscopic parameters
(namely, the particles average values of speeds and polarizations and number  densities). We also assume that this kind of astrophysical medium is composed of three neutrino flavors ($\nu_e$, $\nu_{\mu}$ and $\nu_{\tau}$) forming three independent fluxes moving in the same direction. The generalization of our consideration
for the case when the antineutrino fluxes are also present is straightforward.

Let us now investigate the behavior  of an electron in a dense medium composed of neutrinos
of different flavors. Within a set of assumptions listed above an effective Lagrangian
describing interaction of the electron with fluxes of different flavour neutrinos  is \cite{Studenikin:2004dx,Grigorev:2005sw}

\be
\mathcal{L}=\sum\limits_{i=e,\mu,\tau}\bar{e}\gamma_{\nu}\frac{\epsilon_i (1+\gamma^5)-4\sin^2\theta_W}{2}ef^{\nu}_i,
\label{effective lagrangian_L}
\ee
where $\epsilon_i = -1$ for $i=e$ and $\epsilon_i = +1$ for $i= \mu, \, \tau$. Each of the
flavour neutrino $\nu_{i}$ fluxes is characterized by the neutrino matter potential
\be\label{f}
f^{\mu}_i = G(n_i,n_i\bs{v}_i),
\ee
where the neutrino number  densities $n_{i}$ in the rest frame of the electron  are determined by the corresponding invariant number densities $n^{0}_{i}$ in the rest frame of the particular flavour neutrino flux,
\be\label{n}
n_{i}=\frac{n^{0}_{i}}{\sqrt{1-\bs{v_{i}}^2}},
\ee
$v_{i}$ is the average speed of the flavour neutrino in the flux. Here $G=\frac{G_F}{\sqrt{2}}$, and $G_F$ is the Fermi constant.

We suppose that all flavour neutrino fluxes have the same speed, $\bm {v}_i=\bm {v}$,
and introduce the average value $n$ of the neutrino number densities and the parameter $\delta_e$,
\be \label{n_delta}
n=\frac{n_e+n_\mu+n_\tau}{3},\qquad \delta_{e}=\frac{n_\mu+n_\tau-n_e}{n}.
\ee
Using these notations we  rewrite the effective Lagrangian (\ref{effective lagrangian_L}) in the form,
\be\label{effective lagrangian}
\mathcal{L}=\bar{e}\gamma_{\mu}\frac{c+\delta_e\gamma^5}{2}ef^{\mu},
\ee
where $f^\mu = Gn(1,\bs{v})$ is the effective neutrino potential and  $c=\delta_e-12\sin^2\theta_W$.
For the relativistic neutrinos $v\simeq 1$,  and the effective neutrino potential is
\be\label{fv1}
f^{\mu}=G(n,0,0,n).
\ee
From here we suppose that the neutrino flux propagation is relating to the direction of an axis $z$.

From the Lagrangian (\ref{effective lagrangian}) one gets the following
modified Dirac equation for an electron in the relativistic neutrino flux,
\be\label{modified Dirac equation}
\{\gamma_{\mu}p^{\mu}+\gamma_{\mu}\frac{c+\delta_e\gamma^5}{2}f^{\mu}-m\}\Psi(x)=0,
\ee
here $m$ and $p^{\mu}$   are the electron mass and  momentum.

\section{Exact solution and kinematics}
In the considered case of a relativistic neutrino flux ($v\sim 1$) Eq. (\ref{modified Dirac equation}) with the matter potential given by Eq. (\ref{fv1}) can be solved exactly.
Performing evaluations similarly to those described in \cite{Studenikin:2004dx}, we get
 the energy spectrum of an electron in the presence of the relativistic neutrino
flux (\ref{modified Dirac equation}),
\begin{gather}\label{branch E_s}
E^{\varepsilon}_{s}({p})=\varepsilon\sqrt{m^2+{p}_\bot^2+\Big(p_3+\frac{Gn}{2}\big(c-s\delta\big)\Big)^2}
-\frac{Gn}{2}\big(c-s\delta\big),
\end{gather}
where $\delta$ is the absolute value of $\delta_e$, $p_3$ is the electron momentum in the direction of the neutrino flux propagation and $p=\sqrt{p_{3}^2+ {p}_\bot ^2}$ is the total electron momentum.
The value $\varepsilon=\pm 1$ corresponds to the  positive and
negative frequency solutions (for the electron $\varepsilon = +1$).

The energy spectrum
(\ref{branch E_s}) also contains the second integer number $s=\pm 1$. Comparing Eq. (\ref{branch E_s}) with corresponding spectra of an electron \cite{Grigoriev:2006rv}
or neutrino \cite{Studenikin:2004dx,Grigorev:2005sw} in nonmoving matter
we conclude that the number $s=\pm 1$ distinguishes two possible electron spin states.

\begin{figure}[h!]
\begin{center}
\psset{xunit=0.75 cm, yunit=0.75 cm}  
\begin{pspicture*}(-7,-3)(7,7)  
\psline[linewidth=0.5 pt]{->}(-7,0)(7,0) 
\psline[linewidth=0.5 pt]{->}(0,-1)(0,7) 
\psplot[showpoints=false, linewidth=1.5 pt, plotstyle=curve, algebraic=true]{-1}{6}{2.5+((x-2.5)^2+3)^(1/2)}
\psplot[showpoints=false, linewidth=1.5 pt, plotstyle=curve, algebraic=true]{-6}{6.4}{-1.5+((x+1.5)^2+3)^(1/2)}
\psplot[showpoints=false, linewidth=1 pt, linestyle=dotted, algebraic=true]{-1}{6.5}{x}
\psline[linewidth=1 pt, linestyle=dotted]{-}(-1.5,0)(-1.5,0.2)
\psline[linewidth=1 pt, linestyle=dotted]{-}(2.5,0)(2.5,4.2)
\rput[l](0.1,-0.2){$0$}
\rput[l](0.1,6.8){$E$}
\rput[l](6.5,-0.3){$p_3$}
\rput[l](-3.5,-0.4){$-\frac{Gn}{2}(c+\delta)$}
\rput[l](1.5,-0.4){$-\frac{Gn}{2}(c-\delta)$}
\rput[l](-6,2.5){$E_-$}
\rput[l](-1,5.7){$E_+$}
\rput[l](4,3.5){$E=p_3$}
\end{pspicture*}
\end{center}
\vspace{-1cm}
\caption{
The dependence of the electron energies in two different spin states, $E_+(\bs{p})$ and $E_-(\bs{p})$, on electron momentum component $p_{3}$.}
\label{figure spectrum}
\end{figure}
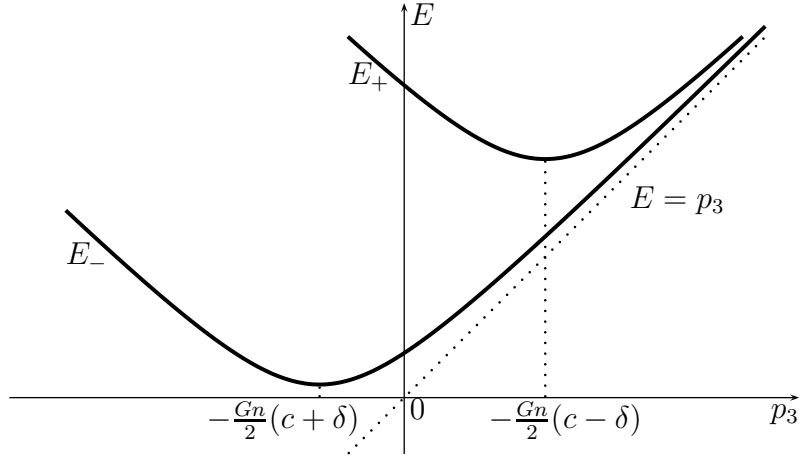

Thus we see that the obtained spectrum branches are classified by
both the frequency sign $\varepsilon=\pm 1$ and the spin sign $s=\pm1$
without explicit introduction and using of the spin operator.
Two particular electron energy branches $E^{\varepsilon}_{s}({p})_{|\varepsilon=+1}=
E_{s}({p})$ with $s=\pm1$ as functions of the momentum $p$ are plotted in Fig. \ref{figure spectrum}.
It is clearly seen that two corresponding curves have no common points that means
there are no energy states with ``undefined'' spin.

It follows from Eq.(\ref{branch E_s}) that for the electron ($\varepsilon=+1$)
there are two different spin states with the same momentum ${p}$. It is interesting to note that
the energy of the state characterized by $s=+1$ always exceeds the energy of the state
with $s=-1$ for the fixed value of the electron momentum $p$.
Indeed, from Eq.(\ref{A_3}) of Appendix we get
\begin{gather}\label{E-E}
E_+({p}) - E_-({p})=\sqrt{m^2+{p}_\bot^2+\Big(p_3+\frac{Gn}{2}\big(c-\delta\big)\Big)^2}-\\-
\sqrt{m^2+{p}_\bot^2+\Big(p_3+\frac{Gn}{2}\big(c+\delta\big)\Big)^2}+Gn\delta>0,
\end{gather}
thus it is always $E_{+}({p})> E_{-}({p})$.

The modified Dirac equation (\ref{modified Dirac equation})
can be solved exactly. For the electron in $s=+1$ spin state the wave function can be found in the following form,
\begin{gather}\label{wave function_i}
\psi_i(\bs{r},t)=\frac{1}{L^{\frac{3}{2}}C_+}e^{i(-E_+t+ \bs{p}\bs{r})}
\begin{pmatrix}
0\\m\\p_\bot\rme^{-i\phi}\\E_{+} -p_3
\end{pmatrix},
\end{gather}
and in $s=-1$ spin state the wave function is
\begin{gather}\label{wave function_j}
\psi_f(\bs{r},t)=\frac{1}{L^{\frac{3}{2}}C_-}e^{i(-E_{-}t+ \bs{p}\bs{r})}
\begin{pmatrix}
E_{-}-p_3\\-p_\bot e^{i\phi}\\m\\0
\end{pmatrix},
\end{gather}
where $L$ is the normalization length, $C_+$ and $C_-$ are normalization coefficients given by
\be\label{C}
C_{\pm}=\sqrt{m^2+p_\bot^2+(E_{\pm}-p_3)^2}.
\ee

\section{Spin light of electron in dense neutrino flux}

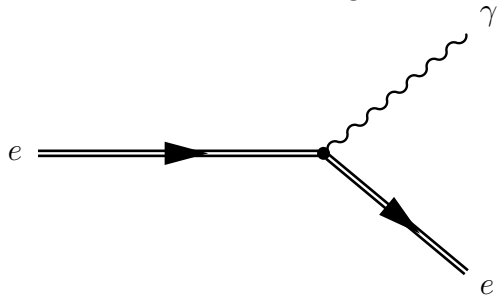
\begin{figure}[h]
\begin{center}
\begin{fmffile}{eLightEmission}
\begin{fmfchar*}(180,90)
  \fmfleft{e_in}
  \fmflabel{$e$}{e_in}
  \fmflabel{$e$}{e_out}
  \fmflabel{$\gamma$}{gamma}
  \fmf{heavy}{e_in,Int1}
  \fmf{heavy}{Int1,e_out}
  \fmf{photon}{Int1,gamma}
  \fmfright{e_out,gamma}
  \fmfdot{Int1}
\end{fmfchar*}
\end{fmffile}
\end{center}
\caption{The Feynman diagram of the radiation by the electron in the neutrino flux.}
\label{Feynman diagram SLe}
\end{figure}

Using the ``method of exact solutions" \cite{Studenikin:2005bq,Studenikin:2007zza,Studenikin:2008qk} we consider the radiative transition of an electron with emission of a photon (plasmon) in the presence of the relativistic flux of neutrinos. This process we term the {\it spin light of electron in dense neutrino flux} ($SLe_\nu$). In the first order of the perturbation theory the Feynman diagram of the process (see Fig. \ref{Feynman diagram SLe}) is the one-photon emission diagram with the bold electron lines denoting the exact account for interaction of electrons with the neutrino background .

The element of $S$-matrix defining the process amplitude
is given by (see also \cite{Grigoriev:2006rv}):
\be\label{amplitude basic}
S^{(\lambda)}_{fi} = -e\sqrt{4\pi}\int\,d^4x \bar{\psi}_f(x)({\bs\gamma} \bs{e}^{(\lambda)}{}^\ast)\frac{\rme^{\rmi kx}}{\sqrt{2\omega L^3}}\psi_i(x),
\ee
where $e$ is the electron charge, $\psi_i(x)$ and  $\psi_f(x)$ are the wave functions of the initial and final
electrons in the background neutrino flux, and $k = (\omega, \bs{k})$ and $\bs{e}^{(\lambda)}$ ($\lambda=1,2$) are the momentum and polarization vectors of the emitted photon. The considered radiative process is a quantum transition between two electron spin states with emission of a photon. It can  proceed because the
condition $E_+ > E_-$ is fulfilled.

Consider the case when the relativistic flux of neutrinos is propagating through a shell of electrons that are at rest. This can be used as a model of real situations peculiar for astrophysical settings.

The electron rest frame in moving background is defined as one where the electron energy $E_+$ gets its minimum.
This means that the electron de Broglie wave group velocity is vanishing, $\frac{\partial E_+}{\partial\bs{p}}=0$.
This condition, accounting for the initial electron energy given by Eq. (\ref{branch E_s}), is provided in case the following relations are fulfilled,
\be
p_3 = -\frac{Gn}{2}(c-\delta),\qquad \bs{p}_\bot = 0.\label{condition of nonmoving electron}
\ee
It can be seen that in the background matter the minimum of the electron energy corresponds to a non-zero value for its momentum \footnote{This phenomenon was discussed in \cite{Chang:1988yn,Pantaleone:1991zr, Pantaleone:1992xh,Studenikin:2004dx}.}. For the case of a supernova environment $\frac{Gn}{m}\sim 10^{-8}$ therefore most naturally electrons can be considered as nonrelativistic particles.

Performing integration over time and the spatial coordinates in Eq. (\ref{amplitude basic}) we recover the $\delta$-functions providing the law of energy-momentum conservation for the considered process and the following relations are straightforward
 (the primed quantities describe the final electron state):
\be\label{e_m_concervation}
\begin{aligned}
E_+ (p)&= E'_- (p')+\omega,\\
-\frac{Gn}{2}(c-\delta)&= p_3'+\omega\cos\theta,\\
0 &= p_\bot' + \omega\sin\theta,
\end{aligned}
\ee
where $\theta$ is the angle between the direction of the $SLe_{\nu}$ radiation and neutrino flux propagation.
\begin{center}
\psset{xunit=1 cm, yunit=1 cm}  
\begin{pspicture*}(-3,-2.5)(3,2.5)  

\psline[linewidth=0.5 pt]{-}(-3,0)(3,0) 
\psline[linewidth=0.5 pt]{-}(0,-2.5)(0,2.5) 

\psline[linewidth=0.5 pt]{-}(-3,0)(-3,0.2) 
\psline[linewidth=0.5 pt]{-}(-2.5,0)(-2.5,0.1) 
\psline[linewidth=0.5 pt]{-}(-2,0)(-2,0.2) 
\psline[linewidth=0.5 pt]{-}(-1.5,0)(-1.5,0.1) 
\psline[linewidth=0.5 pt]{-}(-1,0)(-1,0.2) 
\psline[linewidth=0.5 pt]{-}(-0.5,0)(-0.5,0.1) 
\psline[linewidth=0.5 pt]{-}(0.5,0)(0.5,0.1) 
\psline[linewidth=0.5 pt]{-}(1,0)(1,0.2) 
\psline[linewidth=0.5 pt]{-}(1.5,0)(1.5,0.1) 
\psline[linewidth=0.5 pt]{-}(2,0)(2,0.2) 
\psline[linewidth=0.5 pt]{-}(2.5,0)(2.5,0.1) 
\psline[linewidth=0.5 pt]{-}(3,0)(3,0.2) 

\psline[linewidth=0.5 pt]{-}(0,-2.5)(0.1,-2.5) 
\psline[linewidth=0.5 pt]{-}(0,-2)(0.2,-2) 
\psline[linewidth=0.5 pt]{-}(0,-1.5)(0.1,-1.5) 
\psline[linewidth=0.5 pt]{-}(0,-1)(0.2,-1) 
\psline[linewidth=0.5 pt]{-}(0,-0.5)(0.1,-0.5) 
\psline[linewidth=0.5 pt]{-}(0,0.5)(0.1,0.5) 
\psline[linewidth=0.5 pt]{-}(0,1)(0.2,1) 
\psline[linewidth=0.5 pt]{-}(0,1.5)(0.1,1.5) 
\psline[linewidth=0.5 pt]{-}(0,2)(0.2,2) 
\psline[linewidth=0.5 pt]{-}(0,2.5)(0.1,2.5) 

\psline[linewidth=1 pt]{->}(2.2,0.4)(3,0.4) 

\psplot[showpoints=false, linewidth=1 pt, linestyle = dotted, plotstyle=curve, polarplot=true, algebraic=true]
{0}{TwoPi}{2*1/(1-cos(x)+1)}
\psplot[showpoints=false, linewidth=1 pt, linestyle = dashed, plotstyle=curve, polarplot=true, algebraic=true]
{0}{TwoPi}{2*4/(1-cos(x)+4)}
\psplot[showpoints=false, linewidth=1 pt, plotstyle=curve, polarplot=true, algebraic=true]
{0}{TwoPi}{2*10/(1-cos(x)+10)}
\rput[l](0.1,-0.2){$0$}
\rput[l](2.7,0.6){$\bs{v}$}
\end{pspicture*}
\end{center}
\begin{figure}[h]
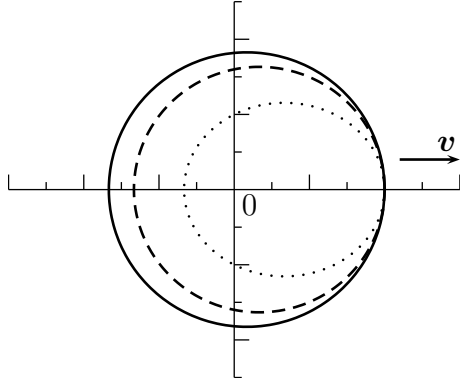

\caption{The angular dependence of the photon energy for different neutrino flux densities $\frac{m}{Gn\delta}=1$ (dotted),
$\frac{m}{Gn\delta}=4$ (dashed) and $\frac{m}{Gn\delta} = 10$ (solid), 
$\bs{v}$ indicates the direction of the neutrino flux.}
\label{figure angular dependence of frequency}
\end{figure}

The emitted photon energy can be obtained as the only solution of Eqs. (\ref{e_m_concervation}),
\be\label{frequency}
\omega = \frac{m}{1-\cos\theta+\frac{m}{Gn\delta}}.
\ee
From Eq. (\ref{frequency}) it follows that in general the emission is possible in all directions. The angular dependence of the photon energy for different neutrino flux densities is shown in
Fig.\ref{figure angular dependence of frequency} . For the most realistic case, when  $Gn\delta\ll m$,
the photon energy simplifies to
\be\label{omega_2}
\omega = Gn\delta.
\ee
Thus, for an initial charged particle with rather large mass or
for the case of the background environment  with enough small density the emitted photon energy
does not depend on the direction of radiation and is determined only by the density of the environment.

Using expressions for the amplitude (\ref{amplitude basic}), the wave functions of the initial (\ref{wave function_i}) and final (\ref{wave function_j}) electrons, and for the photon energy (\ref{frequency})
for the $SLe_{\nu}$ in the neutrino flux rate and power we get,
\begin{align}
\Gamma&=\frac{e^2}{2}m\int\limits_0^\pi\frac{(1-\cos\theta)^2}
{(1-\cos\theta+a)^3}\sin\theta\,d\theta,\label{radiation rate diff}\\
I&=\frac{e^2}{2}m^2\int\limits_0^\pi\frac{(1-\cos\theta)^2}
{(1-\cos\theta+a)^4}\sin\theta\,d\theta.\label{power diff}
\end{align}
where the parameter
\be
a=\frac{m}{Gn\delta}
\ee
is used.
Further integration over the angle $\theta$
gives the closed expressions for the radiation rate
\be\label{radiation rate}
\Gamma=\frac{e^2}{2}m\Big[\frac{2}{a^2}-\frac{2}{a}+\ln(1+\frac{2}{a})\Big],
\ee
and power
\be\label{radiation power}
I=\frac{4}{3}e^2m^2\frac{1}{a^3(a+2)}.
\ee
Much simpler expressions can be obtained in the three limiting cases,
\begin{align*}
a &\gg 1, &\Gamma&=\frac{4}{3}e^2m\Big(\frac{Gn\delta}{m}\Big)^3, &I&=\frac{4}{3}e^2m^2\Big(\frac{Gn\delta}{m}\Big)^4,\\
a &\sim 1, &\Gamma&=\frac{\ln 3}{2}e^2m, &I&=\frac{4}{9}e^2m^2,\\
a &\ll 1, &\Gamma&=e^2m\Big(\frac{Gn\delta}{m}\Big)^2, &I&=\frac{2}{3}e^2m^2\Big(\frac{Gn\delta}{m}\Big)^3.
\end{align*}
For most of astrophysical environments $Gn\ll m$ and the case ($a\gg 1$) is realized.

\begin{figure}[h]
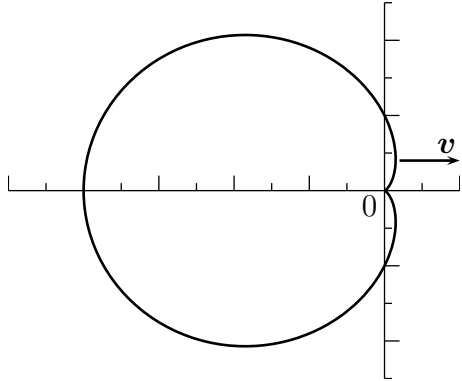

\caption{The angular distribution of the $SLe_{\nu}$ radiation power given by Eq.(\ref{power diff}),
$\bs{v}$ indicates the direction of the neutrino flux.}
\label{figure angular dependence of power}
\end{figure}

From the angular distribution of the $SLe_{\nu}$ given by Eq.(\ref{power diff}) one can estimate the ratio of the radiation power emitted in the forward and backward directions in respect to the neutrino flux propagation,
\be
\frac{I_{forw}}{I_{back}}=\frac{\int\limits_0^{\frac{\pi}{2}}(1-\cos\theta)^2\sin\theta\,d\theta}
{\int\limits_{\frac{\pi}{2}}^{\pi}(1-\cos\theta)^2\sin\theta\,d\theta}=\frac{1}{7}.
\ee
Therefore, a fraction of about  $\frac{1}{8}$ of the total $SLe_{\nu}$ radiation power is emitted in the forward direction (see also Fig. \ref{figure angular dependence of power}).

So that at the space around the supernova core an entire layer (characterized by an overall distance
$R=10 \ km$ from the star centre)
can emit electromagnetic waves with the photon energy decreasing with the increase of
the distance from the centre.

\begin{center}
\psset{xunit=1 cm, yunit=1 cm}  
\begin{pspicture*}(-5,-2.5)(1,2.5)  

\psline[linewidth=0.5 pt]{-}(-5,0)(1,0) 
\psline[linewidth=0.5 pt]{-}(0,-2.5)(0,2.5) 

\psline[linewidth=0.5 pt]{-}(-5,0)(-5,0.2) 
\psline[linewidth=0.5 pt]{-}(-4.5,0)(-4.5,0.1) 
\psline[linewidth=0.5 pt]{-}(-4,0)(-4,0.2) 
\psline[linewidth=0.5 pt]{-}(-3.5,0)(-3.5,0.1) 
\psline[linewidth=0.5 pt]{-}(-3,0)(-3,0.2) 
\psline[linewidth=0.5 pt]{-}(-2.5,0)(-2.5,0.1) 
\psline[linewidth=0.5 pt]{-}(-2,0)(-2,0.2) 
\psline[linewidth=0.5 pt]{-}(-1.5,0)(-1.5,0.1) 
\psline[linewidth=0.5 pt]{-}(-1,0)(-1,0.2) 
\psline[linewidth=0.5 pt]{-}(-0.5,0)(-0.5,0.1) 
\psline[linewidth=0.5 pt]{-}(0.5,0)(0.5,0.1) 
\psline[linewidth=0.5 pt]{-}(1,0)(1,0.2) 

\psline[linewidth=0.5 pt]{-}(0,-2.5)(0.1,-2.5) 
\psline[linewidth=0.5 pt]{-}(0,-2)(0.2,-2) 
\psline[linewidth=0.5 pt]{-}(0,-1.5)(0.1,-1.5) 
\psline[linewidth=0.5 pt]{-}(0,-1)(0.2,-1) 
\psline[linewidth=0.5 pt]{-}(0,-0.5)(0.1,-0.5) 
\psline[linewidth=0.5 pt]{-}(0,0.5)(0.1,0.5) 
\psline[linewidth=0.5 pt]{-}(0,1)(0.2,1) 
\psline[linewidth=0.5 pt]{-}(0,1.5)(0.1,1.5) 
\psline[linewidth=0.5 pt]{-}(0,2)(0.2,2) 
\psline[linewidth=0.5 pt]{-}(0,2.5)(0.1,2.5) 

\psline[linewidth=1 pt]{->}(0.2,0.4)(1,0.4) 

\rput[l](-0.3,-0.2){$0$}
\rput[l](0.7,0.6){$\bs{v}$}

\psplot[showpoints=false, linewidth=1 pt, plotstyle=curve, polarplot=true, algebraic=true]
{0}{TwoPi}{(1-cos(x))^2}
\end{pspicture*}
\end{center}

\section{Polarization properties}

Consider the polarization properties of the discussed $SLe_{\nu}$ radiation. We define two linear polarization vectors $\bs{e}^{(1)}$ and $\bs{e}^{(2)}$ relative to the plane determined by two vectors
$\bs{\kappa} = (\sin\theta\cos\phi, \sin\theta\sin\phi, \cos\theta)$ and $\bs{e}_z=(0,0,1)$.
The vector $\bs{e}^{(1)}$ is perpendicular to this plane and $\bs{e}^{(2)}$ is parallel to it,
\begin{gather}
\bs{e}^{(1)}=\frac{[\bs{\kappa}\times \bs{e}_z]}{\sqrt{1-(\bs{\kappa}\bs{e}_z)^{2}}}= (\sin\phi,-\cos\phi,0),\\
\bs{e}^{(2)}=\frac{[\bs{\kappa}\times \bs{e}^{(1)}]}{\sqrt{1-(\bs{\kappa}\bs{e}_z)^{2}}}=(\cos\theta\cos\phi,\cos\theta\sin\phi,-\sin\theta).
\end{gather}
Decomposing the $SLe_{\nu}$ transition amplitude (\ref{amplitude basic}) in contributions from the photons of the two linear polarizations, determined by the vectors $\bs{e}^{(1)}$ and $\bs{e}^{(2)}$, we get
\be
|S^{(1)}_{fi}| = |S^{(2)}_{fi}|.
\ee
Thus, the $SLe_{\nu}$ radiation rate and power are equally distributed between the two emitted photon linear  polarizations.

\section{Effect of initial electron motion}

The properties of the considered $SLe_{\nu}$ are influenced by possible motion of the initial electron. In particular, the emitted photon energy can be significantly shifted from the value given by Eq.(\ref{frequency}) if the initial electron is moving.

\begin{figure}[h]
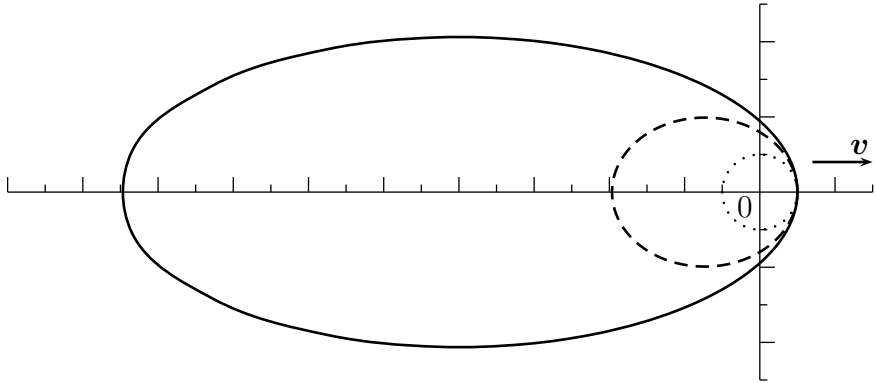

\caption{The photon energy angular dependence for a fixed neutrino flux density $\frac{m}{Gn\delta}=10$ and
different electron momenta:
$\frac{\tilde{p}_3}{m}=0.1$ (dotted line), $\frac{\tilde{p}_3}{m}=1$ (dashed line) and $\frac{\tilde{p}_3}{m}=3$ (solid line),
$\bs{v}$ indicates the direction of the neutrino flux.}
\label{figure angular dependence of frequency relativistic case}
\end{figure}

Consider the case when the initial electron moves with the momentum given by
\be\label{p3}
p_3 = -\frac{Gn}{2}(c-\delta) + \tilde{p}_3,\qquad {p}_\bot = 0,
\ee
here $\tilde{p}_3$ describes the deviation of the third component of the initial electron momentum
$p_3$ from its value (\ref{condition of nonmoving electron}) at rest. It is also supposed that the final electron is at rest. In case the initial electron moves along the neutrino flux propagation (${p}_\bot = 0$) the emitted photon energy is given by  \be\label{omega_p_3}
\omega = \frac{\sqrt{m^2+\tilde{p}_3^2}-\tilde{p}_3}{1-(1+\frac{\tilde{p}_3}{Gn\delta})\cos\theta+\sqrt{(\frac{\tilde{p}_3}{Gn\delta})^2+(\frac{m}{Gn\delta})^2}}.
\ee
Note that the  radiation photon energy in the forward direction ($\theta = 0$) does not depend on $\tilde{p}_3$ and again $\omega$ is given by Eq.(\ref{omega_2}).
On the contrary, the radiation photon energy in the backward direction (when $\theta = \pi$) is strongly dependent on $\tilde{p}_3$
(see Fig. \ref{figure angular dependence of frequency relativistic case}).

In the case the initial electron is moving against the neutrino flux ($\tilde{p}_3 < 0$) we get
\be\label{omega_-_p_3}
\omega = \frac{\sqrt{m^2+\tilde{p}_3^2}+|\tilde{p}_3|}{2-\frac{|\tilde{p}_3|}{Gn\delta}+
\sqrt{(\frac{\tilde{p}_3}{Gn\delta})^2+
(\frac{m}{Gn\delta})^2}}.
\ee
For big enough values of $\tilde{p}_3$, for instance when $|\tilde{p}_3|\gg m$ and also even
$|\tilde{p}_3Gn\delta|\gg m^2$,
there is a reasonable increase of the emitted photon energy in respect to Eq.(\ref{omega_2}),
\be\label{omega_big_p_3}
\omega = 4\Big(\frac{\tilde{p}_3}{m}\Big)^2Gn\delta.
\ee
For example, if  $|\tilde{p}_3|\sim 100 \ m$ then the photon energy is $\omega\sim 4\cdot 10^4 \ Gn \delta$.
\begin{center}
\psset{xunit=1 cm, yunit=1 cm}  
\begin{pspicture*}(-10,-3)(2,3)  

\psline[linewidth=0.5 pt]{-}(-10,0)(1.5,0) 
\psline[linewidth=0.5 pt]{-}(0,-2.5)(0,2.5) 

\psline[linewidth=0.5 pt]{-}(-10,0)(-10,0.2) 
\psline[linewidth=0.5 pt]{-}(-9.5,0)(-9.5,0.1) 
\psline[linewidth=0.5 pt]{-}(-9,0)(-9,0.2) 
\psline[linewidth=0.5 pt]{-}(-8.5,0)(-8.5,0.1) 
\psline[linewidth=0.5 pt]{-}(-8,0)(-8,0.2) 
\psline[linewidth=0.5 pt]{-}(-7.5,0)(-7.5,0.1) 
\psline[linewidth=0.5 pt]{-}(-7,0)(-7,0.2) 
\psline[linewidth=0.5 pt]{-}(-6.5,0)(-6.5,0.1) 
\psline[linewidth=0.5 pt]{-}(-6,0)(-6,0.2) 
\psline[linewidth=0.5 pt]{-}(-5.5,0)(-5.5,0.1) 
\psline[linewidth=0.5 pt]{-}(-5,0)(-5,0.2) 
\psline[linewidth=0.5 pt]{-}(-4.5,0)(-4.5,0.1) 
\psline[linewidth=0.5 pt]{-}(-4,0)(-4,0.2) 
\psline[linewidth=0.5 pt]{-}(-3.5,0)(-3.5,0.1) 
\psline[linewidth=0.5 pt]{-}(-3,0)(-3,0.2) 
\psline[linewidth=0.5 pt]{-}(-2.5,0)(-2.5,0.1) 
\psline[linewidth=0.5 pt]{-}(-2,0)(-2,0.2) 
\psline[linewidth=0.5 pt]{-}(-1.5,0)(-1.5,0.1) 
\psline[linewidth=0.5 pt]{-}(-1,0)(-1,0.2) 
\psline[linewidth=0.5 pt]{-}(-0.5,0)(-0.5,0.1) 
\psline[linewidth=0.5 pt]{-}(0.5,0)(0.5,0.1) 
\psline[linewidth=0.5 pt]{-}(1,0)(1,0.2) 
\psline[linewidth=0.5 pt]{-}(1.5,0)(1.5,0.1) 

\psline[linewidth=0.5 pt]{-}(0,-2.5)(0.1,-2.5) 
\psline[linewidth=0.5 pt]{-}(0,-2)(0.2,-2) 
\psline[linewidth=0.5 pt]{-}(0,-1.5)(0.1,-1.5) 
\psline[linewidth=0.5 pt]{-}(0,-1)(0.2,-1) 
\psline[linewidth=0.5 pt]{-}(0,-0.5)(0.1,-0.5) 
\psline[linewidth=0.5 pt]{-}(0,0.5)(0.1,0.5) 
\psline[linewidth=0.5 pt]{-}(0,1)(0.2,1) 
\psline[linewidth=0.5 pt]{-}(0,1.5)(0.1,1.5) 
\psline[linewidth=0.5 pt]{-}(0,2)(0.2,2) 
\psline[linewidth=0.5 pt]{-}(0,2.5)(0.1,2.5) 

\psline[linewidth=1 pt]{->}(0.7,0.4)(1.5,0.4) 

\rput[l](-0.3,-0.2){$0$}
\rput[l](1.2,0.6){$\bs{v}$}

\psplot[showpoints=false, linewidth=1 pt, linestyle = dotted, plotstyle=curve, polarplot=true, algebraic=true]
{0}{TwoPi}{5*((1+0.01)^(1/2)+0.1)/(1+101^(1/2))}
\psplot[showpoints=false, linewidth=1 pt, linestyle = dashed, plotstyle=curve, polarplot=true, algebraic=true]
{0}{TwoPi}{5*(2^(1/2)+1)/(1+9*cos(x)+10*2^(1/2))}
\psplot[showpoints=false, linewidth=1 pt, plotstyle=curve, polarplot=true, algebraic=true]
{0}{TwoPi}{5*((1+9)^(1/2)+3)/(1+29*cos(x)+10*10^(1/2))}
\end{pspicture*}
\end{center}

\section{Effect of plasma}

It is well know that in general the electromagnetic wave propagation in the background environment is influenced by the plasma effects. For the case of spin light of neutrino in matter these effects have been discussed in details in \cite{Grigorev:2005sw, Kuznetsov:2007ar,Grigoriev:2012pw}. In the above considerations of the $SLe_{\nu}$ an effect of possible nonzero emitted photon mass has been neglected.

Now consider the effect of nonzero emitted photon mass (the plasmon mass $m_\gamma$).  The kinematic condition for the $SLe_{\nu}$ in this case is as follows,
\be\label{op_cond_1}
-\frac{Gn}{2}(c-\delta) + \sqrt{\tilde{p}_3^2+m^2} > -\frac{Gn}{2}(c+\delta) + m + m_\gamma.
\ee
After simplification this conditions reads
\be\label{op_cond_2}
\tilde{p}_3^2 > 2m(m_\gamma - Gn\delta).
\ee
If $m_\gamma < Gn\delta$  the process is opened for arbitrary values $\tilde{p}_3$ of the initial electron including the case when the initial electron is at rest.

In the
nonrelativistic classical plasma $m_\gamma = \omega_p = \sqrt{\frac{4\pi e^2 N_e }{m_e}}$, where $N_{e}$ is the electron number density in the plasme.
Thus, if the electron number density $N_e$ is relatively small the kinematic condition
(\ref{op_cond_2}) is fulfield. In the case $\frac{m_\gamma}{Gn\delta}\ll 1$ the effect of the plasmon mass
is not important. The latter conditions can be realized at a distance of about $R>10\  km$ from a supernova  centre where  $N_e < 10^{19}\,cm^{-3}$.

The Debye screening of electromagnetic waves is another plasma effect that could be important for the $SLe_{\nu}$ radiation propagation in an environment.  Only photons (plasmons) with energy exceeding the plasmon frequency
\be\label{condition}
\omega>\omega_p
\ee
can propagate in the plasma. This sets an upper bound for the electron number density in the cloud,
\be
N_e<\frac{\omega^2 m_e}{4\pi e^2},
\ee
providing possibility for the $SLe_{\nu}$ propagation.
Substituting the corresponding values of $\omega$ one gets $N_e<10^{21}\,cm^{-3}$. Thus, the electron matter with $N_e\sim 10^{19} \ cm^{-3}$ considered above is quite transparent for the $SLe_{\nu}$.

\section{Conclusion}

The supernova phenomenon, that is the most energetic event in the Universe, provides a very promising
environment for  of the proposed new scheme of the spin light  radiation, $SLe_{\nu}$, applications.

Although  there are still opened issues in understanding of the supernova mechanism, the two well confirmed main components of it are the
following \cite{Bethe:1990mw, Janka:2006fh}: (i) the collapse of the core under gravity, (ii) the supernova explosion. The second is of particular interest in the context of the presented studies because it is accompanied by the neutrino burst.
According to the present understanding of the phenomenon (see, for instance, \cite{Kachelriess:2008ze}), practically all gravitational binging energy stored in the star (about 99\%) is emitted in the form of a dense neutrino flux that penetrates surrounding interstellar medium.
Under these conditions the neutrino flux can be considered as a continuum
moving at a speed close to the speed of light.

Consider this flux of the ultra-relativistic neutrinos propagating through medium that contains a reasonable amount of  electrons. In this case the exact solution for the electron energy spectrum is given by Eq.(\ref{branch E_s})
and the spin light of electrons can be emitted in the quantum transition of an electron from the state with the energy $E_+$ to the state characterized by $E'_-$. Note that the transition from the electron state characterized by $s=-1$ to the state with $s=+1$ is not possible for any values of the initial $p$ and final  $p'$ momenta because in this case the energy-momentum conservation equations system, similar to Eq. (\ref{e_m_concervation}), has no nonzero solution.

The emitted photon energy, Eq.(\ref{frequency}), is determined by the effective neutrino number density $n$ and neutrino flux composition given by Eq.(\ref{n_delta}). In order to get the $SLe_{\nu}$ rate and power let us estimate the effective neutrino number density $n$ for  the ultra-relativistic neutrino flux from a supernova.

The neutrino luminosity at the level of $L_{\nu}\sim 10^{53}\ \frac{erg}{s}$ have been predicted by the recent supernova simulations \cite{Liebendorfer:2004aj,Pastor:2002prl}. The same order of magnitude estimations, $L_{\nu}\sim 10^{52} - 10^{53}\ erg \ s^{-1}$, have been obtained \cite{Lunardini:2006ap} for the $\bar{\nu}_e$ luminosity with the average energy $\langle E_{\bar{\nu}_e}\rangle \sim 12 - 14\ MeV$ from the K2 and IBM experimental data.
The neutrino luminosity at a distance $R$ from the center of a supernova is determined by
\be\label{luminosity}
L\sim \langle E\rangle JR^2,
\ee
where $\langle E\rangle$ is the average neutrino energy
in the flux, $J$ is the neutrino flux density. For $\langle E\rangle\sim 10^7 \ eV$ and
$R\sim 10^6 \ cm$ we find $J\sim 10^{45} \ cm^{-2}s^{-1}$.
The neutrino flux density is proportional to the neutrino number density in the flux,
$J=nv$, where $v\sim 1$ for the ultra-relativistic neutrino flux. From this we obtain the neutrino number density in the rest frame of the electron,
\be\label{n_35}
n\sim 10^{35}\,cm^{-3}.
\ee
The same value for $n$ is used in \cite{Dasgupta:2008prd} for the $\bar{\nu}_e$ number density near
the neutrinosphere of the supernova.

Taking into account (\ref{n_35}), we get that $\frac{Gn}{m}\sim 10^{-8}$. Thus,
from Eq. (\ref{omega_2}) it follows for he $SLe_{\nu}$ photon energy
\be
\omega\sim 1 \ eV.
\ee
From  (\ref{radiation rate}) and (\ref{radiation power})
we also find for the $SLe_{\nu}$ rate and power
\be\label{rate_power}
\Gamma\sim 10^{-19}\,eV\sim 10^{-4}\,s^{-1},\qquad I\sim 10^{-7} {eV} {s^{-1}}.
\ee
The corresponding characteristic time of the $SLe_{\nu}$ process is rather big, $\tau\sim 10^4 s$.
It means that $SLe_{\nu}$ from a single electron is hardly observable.

It is interesting to consider a collective effect of the $SLe_{\nu}$ radiation from a cloud of electrons
in space around a supernova core.
 In \cite{Janka:2006fh,Liebendorfer:2004aj} it is shown that there could be
regions with reasonably high electron density. Thus, if the neutrino flux propagates through  such
electron-rich clouds the $SLe_{\nu}$ effect can be increased.

At the distance $R=10 \ km$ from the star center the electron  number density can be of order
$n_e\sim 10^{19} \ cm^{-3}$.
In this case the amount of $SLe_{\nu}$ flashes per second from $1 \ cm^3$ of the electron matter under the influence of a dense neutrino flux     is $N\sim 10^{15} \ cm^{-3} \ s^{-1} $.
For the energy release per one second
of $1 \ cm^3$ of the considered electron-reach matter under the influence of a dense ultra-relativistic neutrino flux we get
\be\label{delta_E}
\frac{\delta E}{\delta t \delta V}\sim N \omega \sim I n_e \sim 10^{13}\ eV \ cm^{-3} \ s^{-1}.
\ee

It is possible to estimate efficiency of the energy transfer from the total neutrino flux to the electromagnetic radiation due to the proposed $SLe_{\nu}$ mechanism. The ``energy content" of the neutrino flux  can be characterized by the product of the mean value of the neutrino energy  and neutrino number density $\langle E\rangle n$. The corresponding energy characteristics of the $SLe_{\nu}$ radiation
is given by $\omega N_{e}$. Thus, for the $SLe_{\nu}$ luminosity we get the following estimation
\be\label{Lum}
L_{SLe}\sim \frac{\omega N_{e}}{\langle E\rangle n}L_{\nu}\sim  10^{31} \ erg \ s^{-1}
\ee
for the following choice of values $L_{\nu}\sim 10^{53}\ \frac{erg}{s}$, $\omega\sim 1 \ eV$,
$N_e \sim 10^{19}\,cm^{-3}$, $n\sim 10^{35}\,cm^{-3}$ and $\langle E\rangle\sim 10^7 \ eV$.

It is interesting to note that the $SLe_{\nu}$ luminosity can be drastically increased in the case when the emitting electrons are moving with relativistic speed against the neutrino flux propagation. As it follows from
Eq.(\ref{omega_big_p_3}), the energy of the $SLe_{\nu}$ photons also increases in this case. We predict that this should have important consequences in astrophysics and for the supernova process in particular.

\section{Acknowledgments}

We are thankful to Alexander Grigoriev, Alexey Lokhov and Alexei Ternov for many fruitful discussions on the considered phenomenon.

\section{Appendix}

Here we show that $E_{+}({p})> E_{-}({p})$ where the energy of the electron in the relativistic neutrino flux ${E_{s}({p})=E^\varepsilon}_{s}({p})_{|\varepsilon=+1}$ is given by Eq. (\ref{branch E_s}).
From Eq. (\ref{branch E_s}) we get the estimation
\begin{gather}\label{est}
\Biggl|\sqrt{m^2+{p}_\bot^2+\Big(p_3+\frac{Gn}{2}\big(c+\delta\big)\Big)^2}-
\sqrt{m^2+{p}_\bot^2+\Big(p_3+\frac{Gn}{2}\big(c-\delta\big)\Big)^2}\Biggr|=\nonumber\\
=\frac{2Gn\delta\big|p_3+\frac{Gn}{2}c\big|}{\sqrt{\Big(m^2+{p}_\bot^2+p_3+
\frac{Gn}{2}\big(c+\delta\big)\Big)^2}+\sqrt{m^2+{p}_\bot^2+
\Big(p_3+\frac{Gn}{2}\big(c-\delta\big)\Big)^2}}<\nonumber\\
<\frac{2Gn\delta\big|p_3+\frac{Gn}{2}c\big|}{\big|p_3+\frac{Gn}{2}c+
\frac{Gn}{2}\delta\big|+\big|p_3+\frac{Gn}{2}c-\frac{Gn}{2}\delta\big|}.
\end{gather}
Taking into account the following simple inequality
\be\label{A_3}
\frac{|x|}{\big|x+|y|\big|+\big|x-|y|\big|}=
\left\{
\begin{array}{rcl}
\frac{1}{2},& x\geq|y|\\
\frac{1}{2}\frac{|x|}{|y|},& -|y|<x<|y|\\
\frac{1}{2},& x\leq-|y|
\end{array}
\right\}\quad\leq\quad\frac{1}{2},
\ee
finally we find that
\be\label{A_4}
\Bigl|\sqrt{m^2+{p}_\bot^2+\Big(p_3+\frac{Gn}{2}\big(c+\delta\big)\Big)^2}-
\sqrt{m^2+{p}_\bot^2+\Big(p_3+\frac{Gn}{2}\big(c-\delta\big)\Big)^2}\Bigr|<Gn,
\ee
and validity of Eq. (\ref{E-E}) is just straightforward.

\bibliographystyle{model1a-num-names} 

\bibliography{BalStu_2014} 

\end{document}